\documentclass{aastex61}

\def\({\left(}
\def\){\right)}
\def\[{\left[}
\def\]{\right]}

\received{July 2, 2017}
\revised{August 27, 2017}
\accepted{August 28, 2017; To appear in ApJ 847:96, 2017 Oct. 1}

\shorttitle{Centrally  Concentrated X-ray Radiation from Corona}
\shortauthors{Liu et al.}

\begin{document}
\title{Centrally  Concentrated X-ray Radiation from an Extended Accreting Corona in Active Galactic Nuclei}
\correspondingauthor{B.F.Liu}
\email{bfliu@nao.cas.cn}
\author{B. F. Liu}
\affil{Key Laboratory of Space Astronomy and Technology, National Astronomical Observatories, Chinese Academy of Sciences, Beijing 100012, China}
\affil{School of Astronomy and Space Science, University of Chinese Academy of Sciences, 19A Yuquan Road, Beijing 100049, China}

\author{Ronald E. Taam}
\affil{Academia Sinica Institute of Astronomy and Astrophysics-TIARA, P.O. Box 23-141, 
Taipei, 10617 Taiwan; taam@asiaa.sinica.edu.tw}
\affil{Department of Physics and Astronomy, Northwestern University,  2131 Tech. Drive, 
Evanston, IL 60208, USA} 
\author{Erlin Qiao}
\affil{Key Laboratory of Space Astronomy and Technology, National Astronomical Observatories, Chinese Academy of Sciences, Beijing 100012, China}
\affil{School of Astronomy and Space Science, University of Chinese Academy of Sciences, 19A Yuquan Road, Beijing 100049, China}

\author{Weimin Yuan}
\affil{Key Laboratory of Space Astronomy and Technology, National Astronomical Observatories, Chinese Academy of Sciences, Beijing 100012, China}
\affil{School of Astronomy and Space Science, University of Chinese Academy of Sciences, 19A Yuquan Road, Beijing 100049, China}

\begin{abstract}
The X-ray emission from bright active galactic nuclei (AGNs)  is believed to originate in a hot corona lying above a cold, 
geometrically thin accretion disk.  A highly concentrated corona located within $\sim10$ gravitational 
radii above the black hole is inferred from observations.  Based on the accretion of  interstellar 
medium/wind, a disk corona model has been proposed in which the corona is 
well coupled to the disk by radiation, thermal conduction, as well as by mass exchange \citep{Liu2015, Qiao2017}.
Such a model avoids artificial energy input to the corona and has been used to interpret the spectral features observed in 
AGN. In this work, it is shown that the  bulk emission size of the corona is very small
for the extended accretion flow in our model. More than 80\% of the hard X-ray power is emitted from a small region 
confined within 10 Schwarzschild radii around a non-spinning black hole, 
which is expected to be even smaller accordingly for a spinning black hole.
Here, the  corona emission is more 
extended at higher Eddington ratios.  The compactness parameter of the corona, $l={L\over R}{\sigma_{\rm 
T}\over m_{\rm e} c^3}$, 
is shown to be in the range of 1-33 for Eddington ratios of 0.02 - 0.1. Combined with the electron temperature 
 in the corona, this indicates that electron--positron pair 
production is not dominant in this regime.  A positive relation between the compactness parameter 
and photon index is also predicted.  By comparing the above model predictions with observational features, we find that the model is in agreement with observations. 
\end{abstract}

\keywords{
accretion, accretion discs
-- black hole physics
-- galaxies: active
-- X-rays:galaxies}


\section{Introduction}
The energetic X-ray emission from bright active galactic nuclei (AGNs) is widely believed to originate in a hot
corona  overlying a thin accretion disk \citep[e.g.][]{Vaiana1978,Haardt1991,Haardt1993,Nakamura1993,Svensson1994,Meyer2000a,
Rozanska2000,Liu2002,Liu2015,Merloni2003}.
The optical--UV emission produced in  the inner accretion disk is Compton up-scattered 
into X-rays by the hot corona. A fraction of this X-ray radiation is directly observed as a power-law component, 
with the remainder illuminating the disk to produce the reflection continuum and emission lines in the spectra. 
However, fundamental questions regarding the formation of the corona and its energizing mechanism, as well as its
size and location are still under debate. 

In recent years, evidence from diverse observational diagnostics points to a very compact corona located within  
10 gravitational radii ($R_g$) of the black hole \citep[e.g.][]{Reis2013,Fabian2015}.
This has been deduced from the rapid variability of the 2-10 keV X-ray emission seen from many AGNs 
\citep[e.g.][]{Done1989} and the X-ray 
spectral timing studies based on reverberation analyses of AGN spectra in the scenario of "lamp post" illumination 
\citep[][]{Fabian2009,DeMarco2011,Kara2013,Reis2013,Cackett2014,Emmanoulopoulos2014,Uttley2014}.
Variability analyses of several lensed quasars also indicate small 
hard X-ray emission regions \citep[e.g.][]{Blackburne2006,Blackburne2011,Blackburne2014,Blackburne2015,
Morgan2008,Morgan2012,Dai2010,Mosquera2013}.
Further evidence for a small physical size of the corona stems from varying obscuration of 
the corona by clouds \citep[][]{Risaliti2011, Sanfrutos2013}. Meanwhile,  the effect of corona geometry on 
the emissivity profile of the broad iron line  has been investigated, from which a comparison of observations and theory 
points to a compact X-ray emitting corona \citep[][]{Wilkins2011,Wilkins2012, Wilkins2016}. Combining the observational 
constraints based on the inferred coronal sizes and electron temperatures, \citet{Fabian2015} studied the radiation 
compactness  and suggested the importance of electron--positron pair production in the corona.  

In contrast to the "lamp post" illumination model, \citet{Gardner2016} suggested the existence of an additional 
energetic emission region lying between the corona and thin disk, which produces radiation in the unobservable 
extreme-ultraviolet (EUV) regime. 
Its high-frequency tail would be seen as the "soft excess" and a low-frequency tail as a "big blue bump". 
Such a model was used to interpret the observed interband time lags where the EUV region serves 
as a reprocessor that is illuminated and heated by the X-ray corona, driving the variability in the 
accretion disk \citep{Gardner2016,Edelson2017}. Although the formation of such a structured component is 
unknown, the hard X-ray emission region/corona is very small.  

On the other hand, a hybrid two-component accretion flow was proposed by \citet{Liu2015} to interpret the 
formation of an energetic corona and a cool, thin disk. Here, matter from the interstellar medium or from stellar winds 
is gravitationally captured by a supermassive black hole \citep[e.g.][]{Chakrabarti1995a,Ho2008}.  
In contrast to the case of low-mass X-ray binaries with a stellar-mass 
black hole, where mass is transferred from a companion star via a Roche lobe overflow, the vertically distributed interstellar medium (or stellar 
wind) tends to form a hot accretion flow, which partially condenses, due to the interaction between the disk and corona, to an underlying cool disk as it flows toward the 
black hole.  Such a model leads to the existence of a strong X-ray emitting corona, which has been used to interpret 
the correlation between the photon index and reflection scaling factor. The best fit is obtained for a corona located above the black hole at a height of 10 Schwarzschild radii \citep{Qiao2017}.  

In this paper, we investigate the bulk emission size of the corona and its dependence on the Eddington ratio in the scenario of 
\citet{Liu2015}. We show that the dominant emission region of the corona is indeed very small, although the hot 
accretion flow is quite extended.  In addition, the compactness parameter is calculated, with which we find the 
model is self consistent since the electron--positron pair production is not important.  In 
\S 2, the basic physics of the model is briefly outlined. The properties of radiation size of the corona 
and of the compactness parameter are presented in \S 3 and comparison with observations is discussed in \S 4. 
Our conclusions are presented in Section 5.

\section{Brief Description of the Model}

We consider a standard accretion disk embedded in a hot accretion flow \citep{Liu2015}. The disk and hot corona 
are well coupled by energy and mass exchange as a consequence of the interaction between the two-phase accretion flows.  
With a constant mass-supply rate the thin disk and corona can achieve a steady state, the form of which 
depends on the properties of the gas supplied to the accretion flow. In low-mass X-ray binaries, gas is transferred 
from a companion star via Roche lobe overflow and is mostly constrained in the orbital plane. Hence,
accretion takes place via a thin disk in the outer region. In the inner region, evaporation is efficient, 
leading to the evacuation and truncation of the thin disk if the mass-supply rate to the disk is low 
\citep[e.g.][]{Liu1999,Meyer2000b}.
On the other hand, if the mass-supply rate to the disk is high, evaporation is not sufficient to 
entirely deplete the disk at any distance. The strong emission from the disk cools the inner corona by 
Compton scattering, leading to a high/soft state with little coronal contribution to the X-ray emission
\citep[e.g.][]{Meyer2000b,MeyerHofmeister2012}.
In contrast, the mass supply for accretion in AGNs results 
from the gravitational capture of matter from stellar winds or the interstellar medium by the central supermassive
black hole. Such gas is not necessarily constrained to lie in the disk plane, and hence it can form a hot 
accretion flow in the outer region.  This hot accretion flow, if encountering any residual thin disk, can heat 
the disk gas and deplete it at low mass-supply rates, similar to the case for a stellar-mass black hole. However, 
for a sufficiently high mass-supply rate, the hot gas condenses partially to the disk, maintaining a thin 
accretion disk and a strong corona close to the black hole. Such a strong corona is essential for the
interpretation of the strong X-ray emission observed in bright AGNs  \citep{Liu2015,Qiao2017}. The 
physical mechanisms involved in the disk corona interaction are similar in both the black hole X-ray binary 
systems (BHXRBs) and AGN. The relative 
strength of the two-phase accretion flow in BHXRBs and AGNs  reflects the differing properties of the mass supply.
A continuous supply of matter to the accreting corona in AGNs  provides for strong X-ray radiation, even if condensation 
of coronal gas is efficient.  Such a model avoids artificial energy input and gas supplement to the corona, 
which are the necessary assumptions in previous models, but which remain to be clarified.

Details of the disk corona interaction processes and relevant equations can be found in \citet{Liu2015} and \citet{Qiao2017}.  
The resultant accretion flows  appear somewhat similar to the two-component model developed by  \citet{Chakrabarti1995b}; 
however, the interaction  processes between the cold and hot flows, and boundary condition in our model significantly 
differ.  Specifically, we adopt a hot, geometrically thick flow  \citep{Narayan1994} fed by the AGN environment, without
an unstable self-gravitating region in the thin disk.  Such an accretion flow partially condenses into a residual disk in 
the innermost radial regions as a consequence of mass and radiative coupling between the hot accretion flow and the cool 
disk \citep{Liu2015}. The steady condensation of a hot accretion flow feeds an inner thin disk, which is different from a 
large disk fed by the Roche lobe overflow process in the stellar-mass black hole context.  The model of 
\citet{Chakrabarti1995b} posits a Keplerian optically thick 
flow in the disk midplane and a sub-Keplerian optically thin flow near the disk surface with the viscosity parameter 
decreasing with disk height. Provided that the viscosity in the hot optically thin halo region is sufficiently low, shock 
waves can form in the halo. On the other hand, in our model, based on a large viscosity parameter which is constant 
throughout the corona, such shock waves are not present. Therefore, our boundary condition (gas supply), the geometry of 
the accretion flow (ADAF-like hot flow and passive disk), and the physics of interaction between the disk and corona 
are significantly different from the model of \citet{Chakrabarti1995b}. 

With the same assumption and method as  that of \citet{Liu2015} and \citet{Qiao2017}, we study the energy balance involving 
thermal conduction and Comptonization, and calculate the coronal temperature, density,  and mass-exchange rate 
between the disk and the corona.  The calculations are iterative because illumination by the corona emission and 
the mass-exchange rate are involved in determining the coronal property. The main computational results for the radiation 
spectrum have been presented in \citet{Liu2015} and \citet{Qiao2017}.  In the following, we investigate the radiation 
compactness and discuss the model predictions with respect to the observations. 

\section{Computational Results} 

The basic model parameters include the mass of a non-rotating black hole, $M$, the Eddington-scaled accretion 
rate, $\dot m$, the viscosity parameter, $\alpha$, and the magnetic field as measured by the ratio of gas pressure 
to total pressure, $\beta$.  In calculating the coronal property,  illumination from the corona contributes to the seed photons in addition to the accretion disk photons for Comptonization.   For this purpose, we adopt a lamp post illumination from a height of 10 Schwarzschild radii ($R_S$) and albedo $a=0.15$.  This height is justified by the comparison of the model prediction with the observed reflection scaling factor \citep{Qiao2017}, and it is roughly consistent with the size of the bulk emission region as shown in the following section.   Given $M=10^8M_\odot$, 
$\alpha=0.3$, and $\beta=0.95$, the emission and spectra for $\dot m=0.02, 0.03, 0.05$, and 0.1 are calculated.  

\subsection{Size of the Dominant Emission Region }

\begin{figure}
\plotone{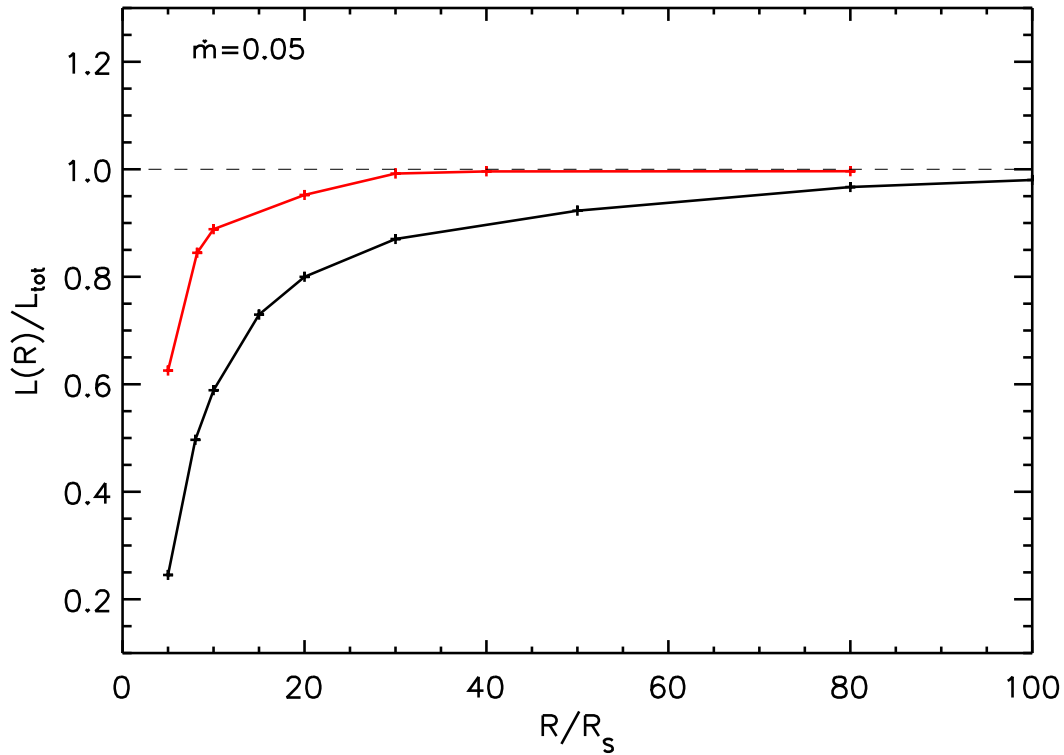}
\caption{\label{f:Lfraction} Cumulative radiation fraction of the total emission (lower curve) and hard X-ray 
emission (upper curve) from the corona as a function of radius in units of the Schwarzschild radius. More than 
$80\%$ of the hard X-ray luminosity is emitted from the inner region within $10R_S$. The difference between hard 
X-ray and total corona radiation indicates that the hard X-ray radiation region is more compact than the 
region emitting soft X-ray radiation.}
\end{figure}

We integrate the coronal radiation power from the innermost stable circular orbit (ISCO) to a distance $R$,  and 
calculate the ratio of this luminosity to the total coronal luminosity, $L(R)/L_{\rm tot}$. Such a ratio 
represents the fraction of cumulative luminosity emitted from the region within $R$, which is used to 
determine the size of coronal emission region. For a  typical accretion rate, $\dot m=0.05$, the 
corona luminosity contributed by different regions as compared with total corona luminosity, $L(R)/L_{\rm tot}$, is 
illustrated in Figure \ref{f:Lfraction}.  For comparison, we also plot the hard X-ray ($2-10keV$) luminosity 
contribution. It can be seen that more than 80\% of the corona emission is contributed from the inner region 
$R\le 20R_S$. The hard X-ray emission is  from a more concentrated region, as more than $80\%$ of the hard X-ray luminosity 
is emitted from the inner region within $10R_S$, and nearly all from within $30R_S$.  This implies that the 
source of hard X-ray radiation is smaller than the soft X-ray radiation as measured by the bulk emission. 
In the following sections,  
the radial distribution of the total corona radiation is used as a probe of the corona radiation size.  
Such an emission size can be converted to  the height of a point-like illumination source above 
the disk within the interpretative framework of lamp post illumination picture (see  Section 4).  Because the hard X-ray emission is contributed from a smaller region as illustrated in Fig.\ref{f:Lfraction} , the size of the bulk emission region calculated from total corona emission is larger than that deduced from hard X-ray reverberation analysis.
 
\begin{figure}
\plottwo{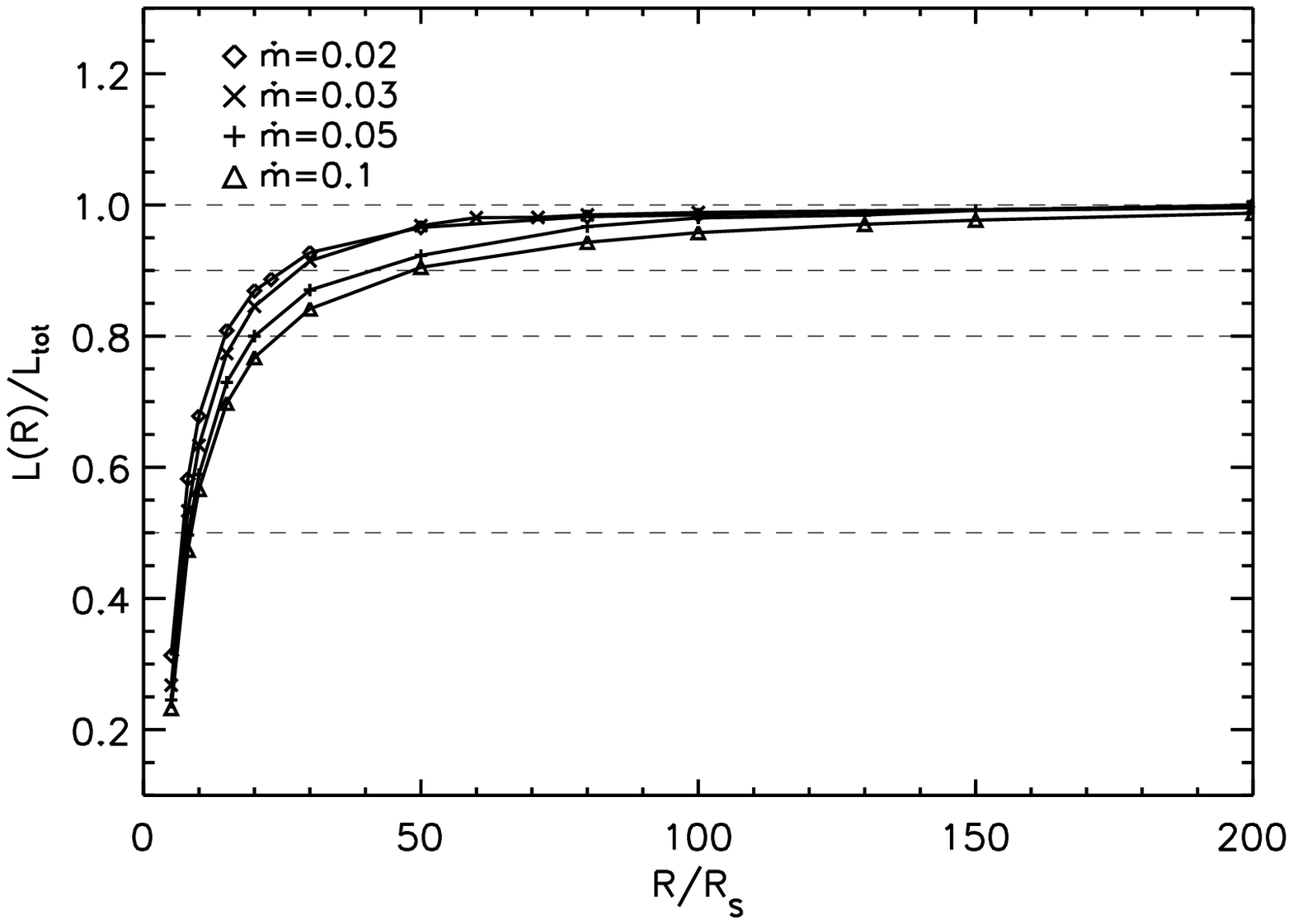}{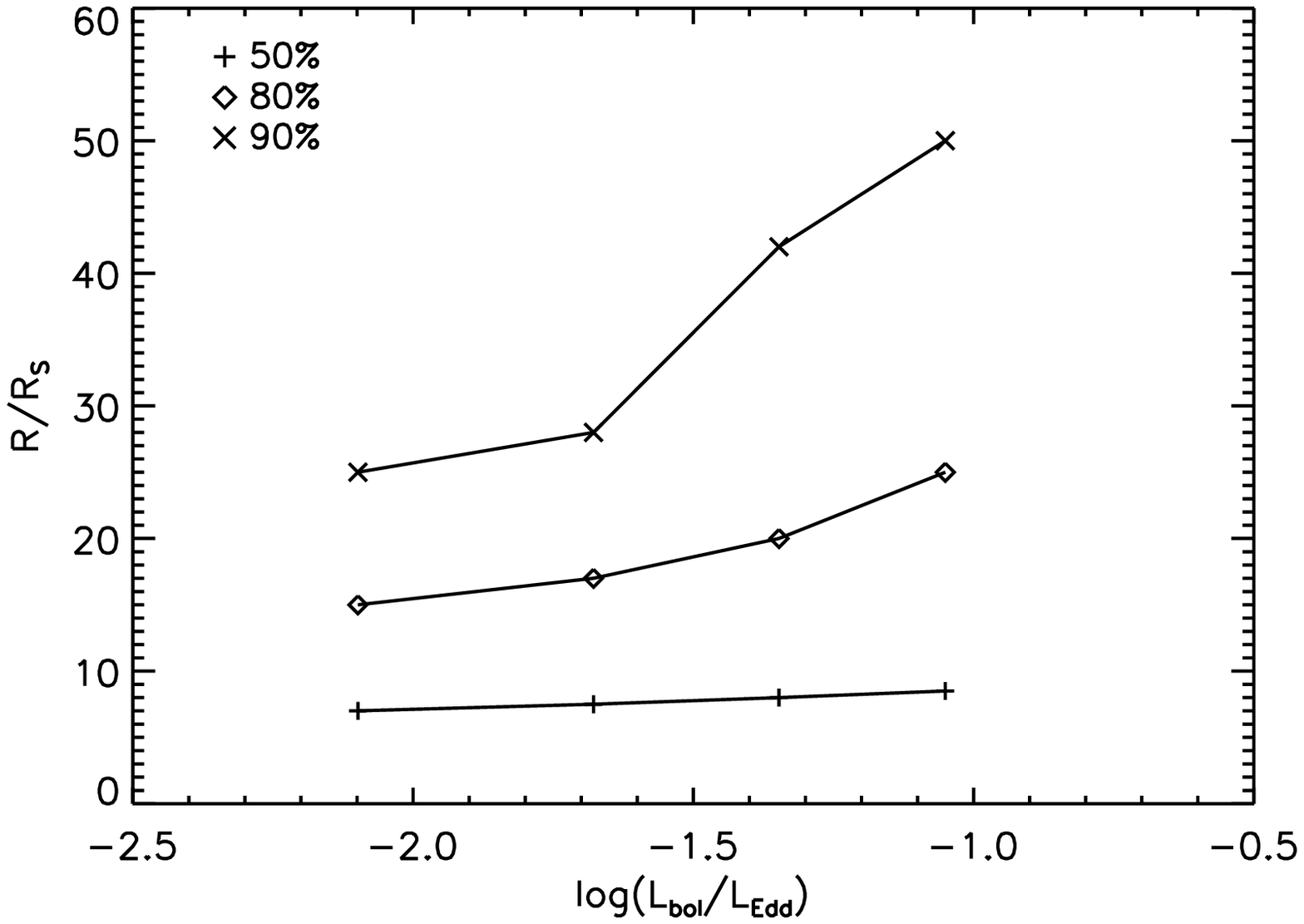}
\caption{\label{f:Lfraction-mdot}Left: the radial distribution of the cumulative radiation fraction of the corona 
for a range of mass accretion rates;  the coronal emission is more extensive at higher accretion rates.  
Right: the size of emitting region as a function of the Eddington ratio.  Fifty percent of the 
coronal luminosity is contributed from a region smaller than $10R_S$ for different Eddington ratios; 80\% of the 
luminosity is emitted from $R<15R_s$ for 
Eddington ratio $\sim 8\times 10^{-3}$. For the same luminosity fraction, the contributed region is larger for a higher 
Eddington ratios. This trend is consistent with observation (e.g. Fig.7 of Kara et al 2016), though the range of our 
Eddington ratio is small.}
\end{figure}

The luminosity fraction from the radial regions is calculated for mass-supply rates of $\dot m=0.02,0.03,
0.05$, and 0.1.  In Fig.\ref{f:Lfraction-mdot}, the distributions for all the accretion rates (left panel) and the size of 
the emitting region as a function of Eddington ratio (right panel) are plotted. It can be seen that the bulk
of the luminosity originates from a more extended region at a higher Eddington ratio (or accretion rate). This result 
can be understood from the radial distribution of the coronal accretion rate for different mass-supply rates, 
as shown in \citet{Liu2015} and \citet{Qiao2017}. For a higher mass-supply rate, the accretion rate in the inner 
corona  increases more steeply with distance as a consequence of the stronger condensation
\citep[see Fig.2 in][]{Qiao2017}.
Therefore, the energy release through accretion is more extended at higher mass-supply rates.   

Within our model assumptions, the above computational results are obtained for a non-spinning black hole.
We note that the size of bulk emission region should be systematically smaller for  a rapidly rotating 
black hole since the accretion flow can extend closer to the event horizon.

\subsection{Radiation Compactness of the Corona}

For a self-consistent examination of our model and  comparison of the model prediction with observations, 
we also calculate the traditional compactness parameter \citep{Guilbert1983}
\begin{equation}\label{e:l}
l={L\over R}{\sigma_{\rm T}\over m_{\rm e} c^3}
\end{equation}
where $R$ is the radius of a spherical source, $L$ is the luminosity produced within the source, $\sigma_T$ is the 
Thomson cross-section, and $m_e$ is the mass of the electron.  This compactness parameter is a measure of the 
importance of coronal cooling as it represents the ratio of the corona cooling time to the light-crossing time. 
If the emitting photons are sufficiently energetic, $h\nu\ga m_ec^2$, photon--photon collisions can lead 
to the creation of electron--positron pairs. The probability of the pair production  can also be measured by the 
compactness parameter.  If $l$ reaches a certain value, a significant fraction of the source luminosity is 
lost in pairs and is then reradiated at lower energies, which can play a major role in determining the 
outgoing spectrum and overall composition of the corona \citep[e.g.][]{Svensson1984}.

For an optically thin corona, the radiation luminosity $L$ is approximately given by the sum of emission 
from the entire region,
\begin{equation}\label{e:Luminosity}
L={4\over 3}\pi R^3 q_{\rm c} ,
\end{equation}
where $q_{\rm c}$ is the radiation cooling rate per unit volume in the corona, i.e., the emissivity of the corona. 
The cooling time of the corona can then be expressed as
\begin{equation}\label{e:t_c1}
t_c={n_{\rm e}\gamma m_{\rm e} c^2\over q_{\rm c} } ,
\end{equation}
where $\gamma=1/\sqrt{1-v^2/c^2}$ the Lorentz factor and $n_e$ the number density of the electrons. Combining 
Eqs.(\ref{e:l})--(\ref{e:t_c1}) and using the definition of the scattering optical 
depth $\tau=n_e \sigma_T R$, we derive the cooling time of the corona compared to light-crossing time as  
\begin{equation}\label{e:t_c2}
{t_c \over R/c}={{4\over 3}\pi \gamma \tau \over l},
\end{equation}
The above expression (Eq.\ref{e:t_c2}) is independent of the radiation mechanism
and is valid for an optically thin corona. The detailed radiation processes 
are implicitly included in the compactness parameter $l$ through the luminosity. Eq.(\ref{e:t_c2}) reveals the 
physical meaning of compactness parameter $l$ as follows.  For a corona with typical value $\tau\sim 0.5$, if 
the electrons are non-relativistic, $\gamma\approx1$,   an electron is cooled by radiation in a light-crossing 
time when $l\approx 2$. A larger value of $l$ indicates the more rapid cooling of electrons with its timescale 
shorter than the light-crossing time.

Note that Eq.(\ref{e:t_c2}) does not imply that radiative cooling is slow for energetic electrons or for high-scattering 
depth.  In fact, the radiation power of a relativistic electron is proportional to $\gamma^2$, independent
of whether the emission is from inverse Compton scattering or synchrotron radiation. Energetic electrons produce 
stronger radiation, $l\propto L\propto \gamma^2$, thereby leading to a shorter cooling time as compared to a light-crossing time, ${t_c\over R/c}\propto 1/\gamma$. This effect is significant for ultra-relativistic electrons.  The 
scattering optical depth also affects the radiation compactness. A large scattering depth in a given region 
(fixed $R$) corresponds to a high-electron density, leading to a high total radiation power by either synchrotron 
or Compton radiation. Thus, the radiation is more compact for larger $\tau$. For ultra-relativistic electrons this 
effect is not as significant as the Lorentz factor $\gamma$, as  $\tau$ is not much less than 1 in the optically 
thin corona, as required by the observed bright X-ray emission in AGN.  The emission power from synchrotron 
radiation or inverse Compton scattering of external photons is proportional to $n_e$, which cancels $n_e$ 
contained in $\tau$ in Eq.(\ref{e:t_c2}). Thus, the ratio of cooling time to light-crossing time is independent 
of the electron density. For the Synchrotron Self Compton (SSC) process, where electrons take part in both 
the synchrotron process and Compton scattering, the emission power is $\propto n_e^2$.  For Bremsstrahlung 
radiation, the radiation power is similar to SSC, $\propto n_e n_i \propto n_e^2$. In these cases, we have ${t_c\over 
R/c}\propto 1/\tau$, indicating a longer cooling time compared with the light-crossing time for lower densities. Such 
a case usually occurs at low Eddington-scaled accretion rates where external Compton scattering is 
unimportant compared to SSC or Bremsstrahlung in an optically thin corona or ADAF.

To summarize, the radiation compactness parameter, $l$, is very large for relativistic electrons. For a thermal 
corona, this signifies a high temperature. With the decrease of temperature, electrons are non-relativistic and the 
radiation compactness parameter is smaller.
 The compactness in a thermal corona also depends on the scattering depth, which is 
reflected in the Eddington-scaled accretion rate in the corona.  At a high accretion rate, the radiation is 
strong and $l$ is large.  When the accretion rate is very low, and SSC or Bremsstrahlung is the dominant 
cooling process, the corona radiation is  no longer strong and hence $l$ is small, as that in an ADAF.
 
\begin{figure}
\plottwo{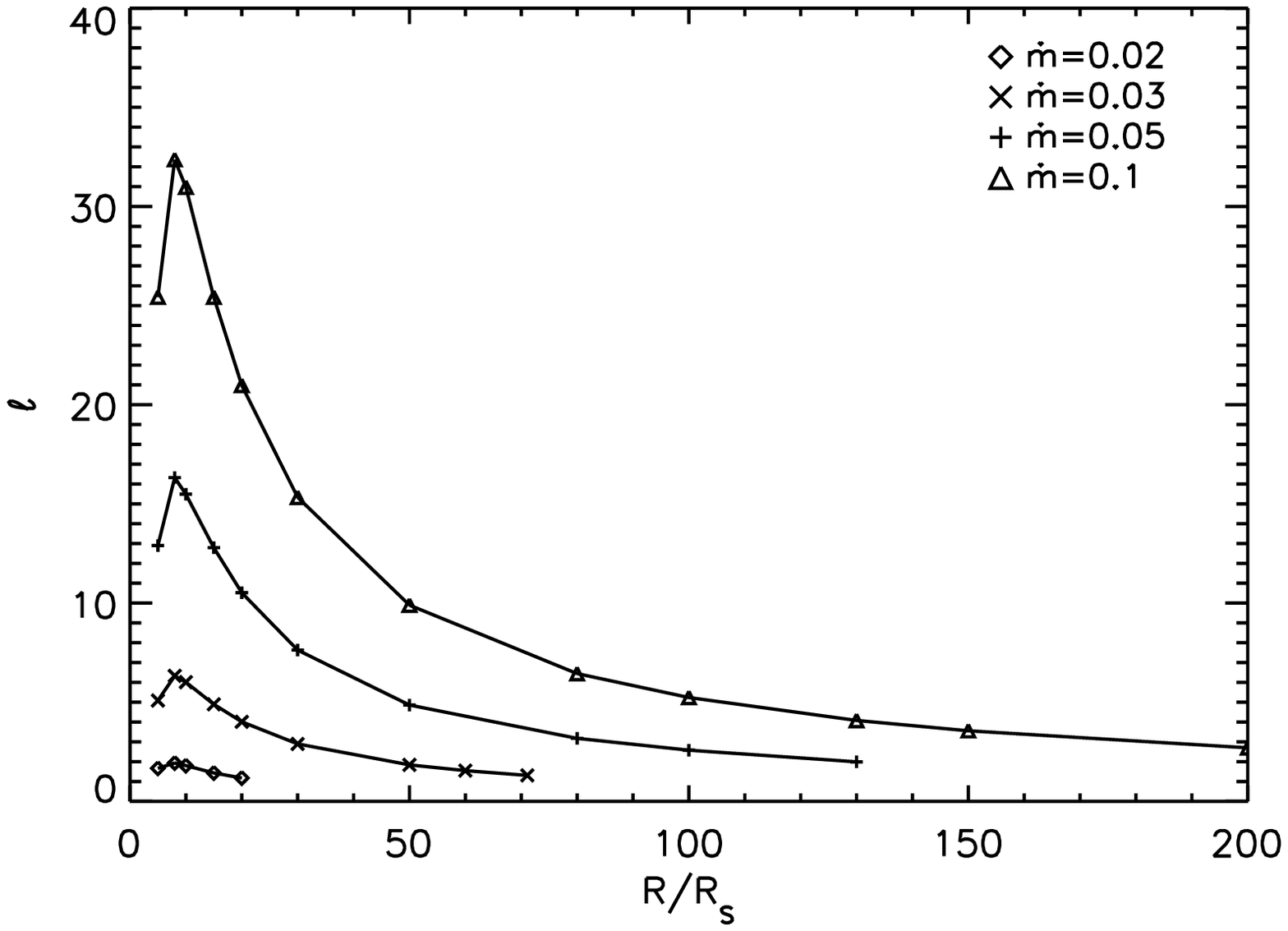}{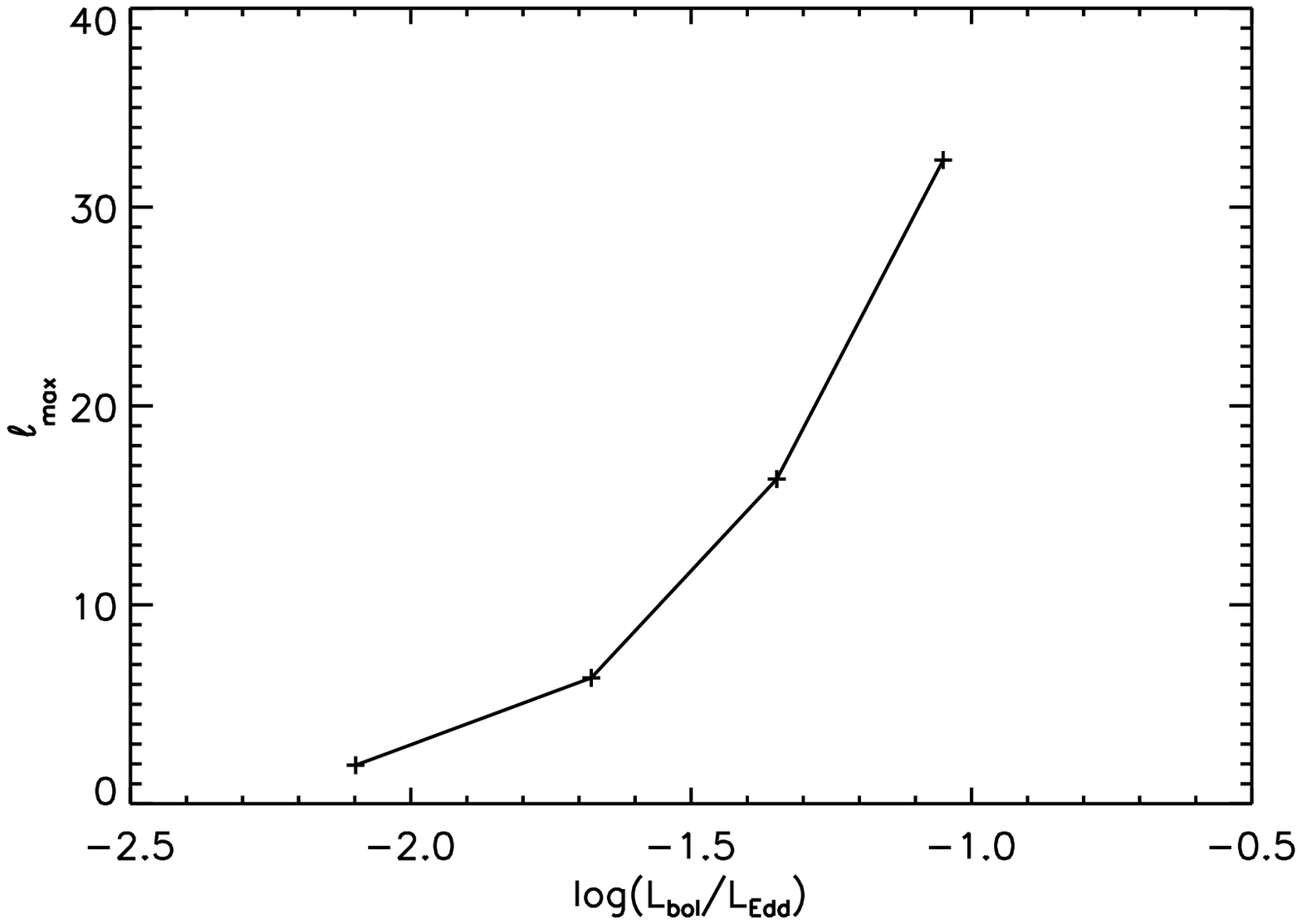}
\caption{\label{f:l-r} Left: the radial distribution of  the compactness parameter $l$ for mass-supply  rates
$\dot m=0.02,0.03,0.05, 0.1$. Right: the dependence of the maximal compactness parameter $l_{\rm max}$ on Eddington ratio.}
\end{figure}

We calculate the compactness parameter of the corona and plot $l-r$ in Fig. \ref{f:l-r} for the mass-supply rates 
corresponding to $\dot m=0.02,0.03,0.05$, and 0.1. It can be seen that the 
compactness parameter increases with decreasing radius
as the flux from the seed photons increases, reaching its maximum at about $8 R_S$ and then
decreasing again as a combination of the Compton {\it y}-parameter and  the density of seed photons
near the ISCO decreases. The value of $l$ ranges from 1 to 33 and is less than 10 in the outer region with $R>50R_S$ 
in the range of the calculated accretion rates.  Taking the maximum compactness parameter as an estimate of the 
compactness of the inner region for each  accretion rate, in Figure \ref{f:l-r} we plot the value of $l_{\rm max}$ versus the
Eddington ratio, where the Eddington ratio is calculated from the spectral energy distribution for a given accretion 
rate. This figure demonstrates that the coronal compactness parameter is larger for higher Eddington ratios, 
indicating stronger radiation in  the corona. 

At very high compactness, photon--photon collisions are efficient, leading to electron--positron pair production if a 
large fraction of the coronal emission is in the high-energy band, $h\nu\sim m_ec^2$. In an optically thin 
thermal corona, the inverse Compton scattering of the soft photons produces a power-law radiation spectrum extending 
to a Wien tail at energies $h\nu\sim 2kT_e$ \citep{Petrucci2001}, which is $3kT_e$ in the case of saturated 
scattering \citep{Rybicki1979}.  For the disk corona model considered here, the electron temperatures 
are $\sim 10^9$K, corresponding to a high-energy cutoff $h\nu \sim 0.3 m_ec^2$. This suggests that the 
high-energy $\gamma$-ray photons are only a small fraction of the total coronal emission. Even if the model 
parameters are adjusted to produce a flat spectrum in the energy range 2-10keV, the number density of 
$\gamma$-ray photons cannot be very high in the thermal corona. This is in contrast to the case for 
non-thermal emission, in which a large fraction of $\gamma$-rays could be expected; in particular, in the case 
where a flat X-ray spectrum 
is observed. 

As the coronal temperature sets an upper limit for the radiation tail of the corona, the probability of pair 
production caused by collisions of high-energy photons depends on the electron temperature.  Detailed radiative 
transfer calculations including both the energy and pair balance yield a critical value of $l$ as a function of 
electron temperature, which depends on the geometry \citep{Svensson1984,Svensson1996,Stern1995}. The most stringent 
constraint to the upper limit for $l$ in the "pair-free" regime is for slab geometry \citep[see Fig.3 in][]{Svensson1996}. 
Our calculations, based on the assumption that radiation originates from inverse Compton scattering, synchrotron, 
and bremsstrahlung processes, yield values of $l$ marginally below the critical value given by 
\citet{Stern1995} or \citet{Svensson1996}. At higher accretion rates beyond our calculations, we expect  stronger coronal radiation, which leads to 
a larger compactness parameter $l$. However, the electron temperature should slightly decrease as a consequence 
of more efficient Compton cooling, which alleviates the importance of
pair production. Hence, pair production is not expected to significantly affect 
the emergent radiation, indicating that our model is self consistent.   

\section{Discussion}
    
X-ray reverberation provides a new technique for investigating the geometry of the accretion flow in the inner region 
of an accretion disk surrounding a supermassive black hole. Observations using the {\it XMM-Newton} and {\it NuSTAR} 
satellites have shown that the soft X-ray excess 
\citep{Fabian2009,DeMarco2013,Reis2013}, broad iron {\it K} line \citep{Zoghbi2012,Kara2013,Kara2016}, 
and Compton hump \citep{Zoghbi2014,Kara2015} all lag the continuum emission. 
This time lag is assumed to be caused by the light-travel time between the corona and the ionized accretion disk, 
providing a probe of the height and extent of the coronal source.  Along this line of investigation, significant efforts 
have been directed toward observationally constraining the corona geometry. It is found that the coronal illumination 
may originate from a height less than $10R_g$ \citep[Detailed data analyses on the reverberation lag and deduced 
height of the corona can be found in][]{Fabian2009,Fabian2015,DeMarco2013,Kara2013,Kara2015,Kara2016,Reis2013}.

As a hot accretion flow, the vertical extent of the corona 
at any distance is approximately equal to the radius. If the illumination from the corona is 
approximated as a point source, as adopted in phenomenological models, the average height of illumination should be less than half of the 
radial extent ($R_{\rm out}$) of bulk emission region,  as estimated from 
\begin{equation}\label{e:averageH}
\langle H\rangle={1\over L}\int_{\rm R_{ ISCO}}^{\rm R_{out}} {H\over 2} {dL\over dR} dR={1\over 4}({\rm R_{out}}+{\rm R_{ ISCO}})<{1\over 2 } {\rm R_{out}},
\end{equation}
where homogeneous radiation in coronal annuli (i.e.${dL\over dR}$=constant) and $H=R$ are assumed.  Taking into account of 
the radially decreasing coronal luminosity of  ${dL\over dR}$, the location of an  illumination 
source should be  smaller than the approximation given in Eq.(\ref{e:averageH}). In \S 3, it was demonstrated that the coronal radiation is, indeed, 
concentrated in the inner region. For example, the radial extent is $\sim 10R_S$ and $\sim 5 R_S$ corresponding to 80\% 
and 60\% of the hard X-ray luminosity respectively for $\dot m=0.05$, implying the bulk emission is respectively  from a region smaller than $\sim 5R_S$ and $\sim 2.5 R_S$.  This suggests that the disk corona model adopted here is roughly consistent with the observations.    

\citet{Kara2016} carried out a survey of the X-ray time lags in a large sample of Seyfert galaxies observed 
with {\it XMM-Newton} and found that the coronal source height tends to be larger for sources at higher Eddington 
ratios, though there is large scatter (in their Fig. 7 and Fig. 8). This is consistent with the theoretical expectation 
that  the size of the 
bulk emission region is a function of the Eddington ratio (right panel in Figure \ref{f:Lfraction-mdot}). 
 
Additional observational support for this correlation is provided by the work of \citet{Gallo2015} on the 
narrow-line Seyfert galaxy Mrk 335.  Here, deep X-ray observations of this galaxy in the low-flux state when 
combined with observations of the source when it was in a state 10 times brighter showed that the scenario of blurred 
reflection from an accretion disk \citep[e.g.][]{Ross2005} is more likely than the partial-covering absorption 
scenario \citep[e.g.][]{Tanaka2004}  based on both long-term and rapid variability.  Adopting the reflection model, 
\citet{Gallo2015} found that the high-flux state of Mrk 335 is consistent with continuum-dominated, jet-like 
emission, of which the ejecta is confined and bound to within $\sim 25R_g$. On the other hand, the corona 
becomes compact, extending to only $5 R_g$ from the black hole during the low flux state, with its spectrum 
dominated by the reflection component.  
\citet{Keek2016} performed a multi-epoch spectral analysis of all {\it XMM-Newton, Suzaku,} and  {\it NuSTAR} 
observations of Mrk 335. The low-flux state was interpreted as a compact and optically thick corona, which is 
located close to the inner disk, whereas in the high-flux state the corona is optically thin and extends vertically 
further away from the disk.  Although the presumed coronal geometry differs from our model,  these results point to 
the same conclusion that the corona is more extended at a higher Eddington ratio.
  
The above observational features can be understood as a consequence of the disk corona interaction. At low 
mass-supply rates, the coronal gas can remain in the corona without significant condensation.  Thus, the 
radial distribution of the accretion rate in the corona is flat. The corona radiation is concentrated in the 
innermost region where there is a greater release of radiation than in the outer region. This innermost 
region corresponds to a very small height ($H<R/2 $) when it is approximated as a point source. At a high mass 
supply rate, gas condenses to the disk, leading to a stronger disk.  This, in turn, boosts coronal cooling 
through external Compton scattering and, hence, enhances gas condensation in the inner region, resulting in 
a steep radial distribution of the accretion rate in the corona and extended radiation.  The illumination 
flux to the innermost disk (where most of the blurred Fe lines are produced) depends on the location of the 
illumination source and is given by
\begin{equation}
F_{\rm il} = { L_{\rm il} \over 4\pi}{H\over {(R^2+H^2)}^{3/2}}\propto { L_{\rm il} \over R^2} ,
\end{equation}
where $L_{\rm il}$ is the luminosity of a point source located at a radius of $R$ and a height  of $H\sim R$.
Hence, the radiation from the innermost corona plays a key role in affecting the reflection component, 
 as a consequence of the distance-dependent illumination flux.  
The strength of the reflection relative to the direct emission  increases with the fraction of luminosity 
from the innermost region of the corona.  As shown in Figure 2, the luminosity is 
 distributed more extensively at a high Eddington ratio, leading to the expectation that reflection is less 
important. In this sense, it is not necessary to introduce an outflowing corona at high Eddington ratios.  
The change in the geometry of the corona can be interpreted as a smooth variation in the hot corona and the thin 
disk as a function of the mass-supply rate to the accretion flow.  More detailed computations with a radially 
distributed illumination in the scenario of the condensation model is required for quantitative comparison 
with observations. 
 
The illumination of a disk from a corona for various geometries (i.e., a point source, a radially extended slab 
corona, a vertically extended cylindrical corona, and a collimated outflow) has been investigated to determine the 
X-ray reflection emissivity profiles based on general relativistic ray-tracing simulations
\citep{Wilkins2011,Wilkins2012,Wilkins2016}.
Combined with the observed reverberation time lags, \citet{Wilkins2012}  fitted 
the broadened iron {\it K} emission line in the spectrum of 1H 0707-495 and found an extended region of primary X-ray 
emission located as low as $2R_g$ above the accretion disk, extending outwards to a radius of around $30R_g$.  Although 
the geometry is simple and the extended corona is homogeneous, these investigations provide a constraint on the 
geometry of the illumination source. Given that the corona emission is inhomogeneous and the  scale height 
varies with distance in our model, the X-ray reflection emissivity is expected to differ from these simplified models. 

\citet{Kara2015,Kara2016}  point out that the true light-travel time between the corona and 
the disk is longer than the observed lag measured by X-ray reverberation due to dilution effects.  The direct 
emission from the corona, as our model reveals, unavoidably contributes a fraction to the reflection band.  
This part, as the leading variation component, will dilute the lag of  the reflected component. When converting the 
observed lag into a distance, the correction factor associated with the dilution effect is dependent on the 
relative contribution of the direct emission in the reflection band and can lead to an upward revision of the 
distance by a factor up to 4 \citep{Kara2015}. 
For example, the dilution to the reflection component in the soft X-ray band ($<1keV$) is stronger than that 
for the iron $K$ line band.  As a consequence, the true size of the corona could be a few times larger than the values 
derived from reverberation. By taking the dilution effect into account, the radiation size deduced from the 
observations would be in better agreement with that predicted by the corona condensation model. 

In  addition to constraints imposed by variability, micro-lensing is another potential tool for estimating 
the emission size of the corona.  
Observations of lensed quasars have shown that the half-light radius of the hard X-ray emission region is $20R_g$ 
for PG 1115+080 \citep{Morgan2008}, $\la 16R_g$ for QJ 0158-4325 \citep{Morgan2012}, $\approx 20R_g$ for Q 2237+0305 
\citep{Mosquera2013}, $\sim 10R_g$ for HE 0435-1223 \citep{Blackburne2014}, $25R_g$ for HE 1104-1805
\citep{Blackburne2015}, and $<24R_g$ for J 0924+0219 \citep{MacLeod2015}.  These measurements, 
as listed in Table 1, constrain the quasar X-ray 
emission size to be $\sim 20 R_g$ \citep[see also][]{Blackburne2006,Pooley2006,Chartas2009,Dai2010}.
\citet{Chen2011} and \citet{Mosquera2013} detected energy-dependent X-ray micro-lensing in Q 2237+0305, 
suggesting that the hard X-ray component is more compact than the soft one.  The size of bulk X-ray emission predicted 
by the condensation model is roughly consistent with micro-lensing observations, as shown in Fig.\ref{f:Lfraction}  
where it can be seen that the hard X-ray component is more compact than the soft X-ray component.

\begin{table}[ht]
\caption{The Half-light Radius of the Hard X-ray Emission Region for Micro-lensed AGN}
\centering                          
\begin{tabular}{l| c c c|c}\hline\hline
 Source & $\log R$(cm) & $M/M_\odot$ & Reference & $R/R_g$\\\hline
 PG 1115+080 & $15.6_{-0.9}^{+0.6}$& $1.2\times 10^9$&Morgan et al. 2008 &20\\
 Q J0158-4325 &$\la 14.6$&$1.6\times 10^8$&Morgan et al. 2012&$\la 16$\\
 Q 2237+0305 &15.46&$\approx 10^9$&Mosquera et al. 2013&$\approx 20$\\
 HE 0435-1223&$<14.8$&$\ga 5\times 10^8$&Blackburne et al. 2014&$\sim 10$\\
 HE 1104-1805&15.33&$5.9\times 10^8$&Blackburne et al. 2015&25\\
 J 0924+0219&$<15$&$2.8\times 10^8$&MacLeod et al. 2015&$<24$\\\hline 
 \end{tabular}
\end{table}

As remarked in \S 3, the value of the compactness parameter $l$ measures the importance of radiation cooling on 
the light-crossing time, revealing the importance of the pair production process operating in the corona. The value 
predicted by our model ranges from 1 to 33, which indicates strong emission in the inner corona.  The electron 
temperatures in the region with maximal compactness parameter are $\sim 10^9$K.  Thus, the value of $l$ is marginally 
below the critical value for the pair production to 
affect the spectrum, suggesting that the inverse Compton scattering, synchrotron and bremsstrahlung emission can 
be regarded as the dominant cooling processes in our model.

This differs somewhat from the conclusion of \citet{Fabian2015}, who examined the compactness parameter $l$ for a collection of 
objects.  In their study, the coronal sizes were based on the modeling of the reflection spectrum, measurements 
by X-ray reverberation or by micro-lensing analysis.  If no measurement existed a value of $10R_g$ was adopted. The derived 
compactness $l$ values are mostly below 100, but are systematically larger than the values predicted by the condensation 
model. If the dilution effect in reverberation measurements were taken into account, they can be expected to be more 
consistent.  On the other hand, \citet{Fabian2015} point out that the general relativistic effects associated with the 
gravitational redshift and light-bending can boost the intrinsic values of $l$ by factors of around 2-10 above the 
observed estimates.  However, such an effect is predicated under the assumption of an extremely small corona. If the 
corona size/height is not as small as 10$R_g$, as our model predicts, or as observationally inferred with dilution 
taken into account, the contribution due to general relativistic effects could be overestimated.  
Additionally, the radial extent of the corona is larger than the vertical height deduced from reverberation lags, 
as discussed above. If this effect is 
considered, the value of $l$ in \citet{Fabian2015} is closer to that predicted by the disk corona model.  This 
suggests that the pair production may not be the dominant process in cooling and heating of the corona, though it 
may be involved. 

\begin{figure}
\plotone{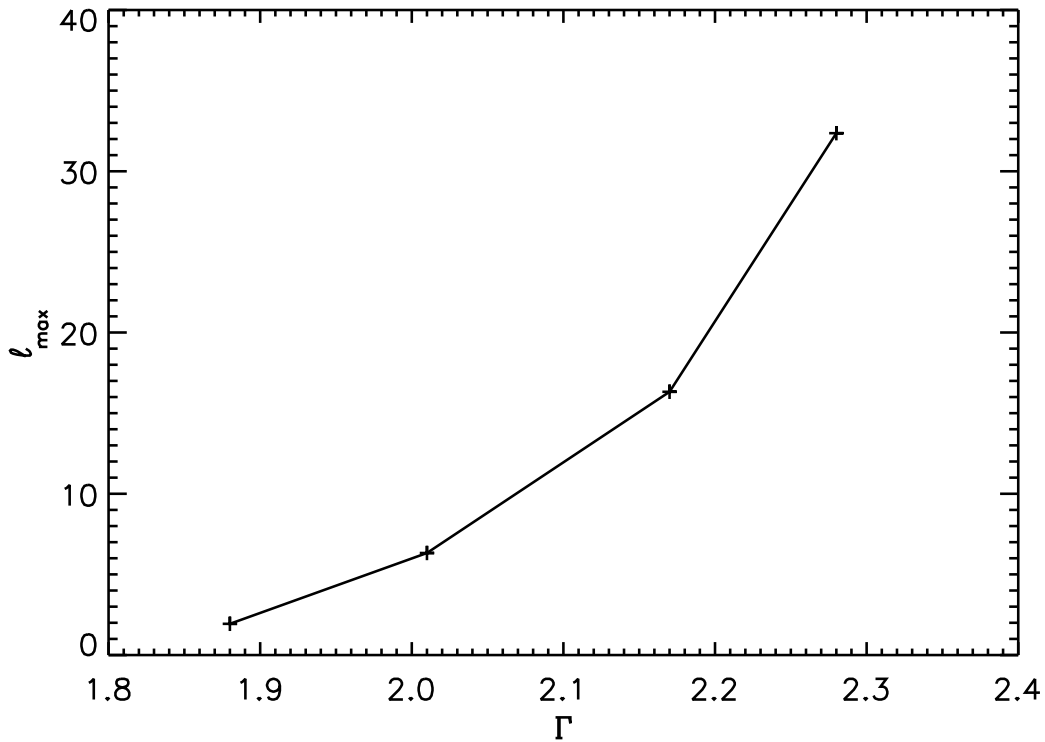}
\caption{\label{f:l-Gamma}  Dependence of the maximal compactness parameter $l_{\rm max}$ on photon index $\Gamma$ in 2-10keV energy band. }
\end{figure}

To compare with observation, we illustrate the relation between the maximal compactness parameter ($l_{\rm max}$) 
and the 2-10keV photon index ($\Gamma$)  in Fig.\ref{f:l-Gamma}. As both 
the compactness parameter  and the photon index increase with Eddington ratio, a positive correlation between 
these properties is expected, as shown in Figure \ref{f:l-Gamma}. Although the the curve of $l_{\rm max}-\Gamma$ 
only covers a small range in both the compactness parameter and photon index, it is close to the curve derived 
for a slab corona and below the curve for a hemispheric corona in energy and pair balance \citep{Stern1995}.  
However, the compactness parameter for given $\Gamma$ is systematically smaller than the data shown in Fig.8 of 
\citet{Fabian2015}, which could be caused by different size of radiation region as discussed above.

\section{Conclusion}

The bulk emission size of the corona and its radiation compactness are investigated on the basis of the 
corona condensation  model \citep{Liu2015,Qiao2017}. It is found that the emission region is small in spite 
of a radially extended corona flow. The hard X-ray emission is contributed by an even smaller region  than  that of the soft X-rays, 
as measured by the bulk emission from the corona.  Specifically,  more than 80\% of the hard X-ray luminosity  is emitted from a region less than $10R_S$, comparing favorably with micro-lensing measurements. 
Combining the radial distribution of coronal emission and the assumption of a geometrically thick corona in the model, we show that the height of a point-like illumination source in the lamp post model should be smaller than half of the radial size of the bulk emission.  This implies that the illumination height  is roughly in  agreement with that deduced from reverberation time lag measurements.   
At lower accretion rates/Eddington ratios, the model predicts an even smaller  region of bulk  emission, which  is also consistent with the purported correlation found in the observations.

The traditional compactness parameter has been calculated to estimate the radiation cooling timescale as 
compared to the light-crossing timescale. The compactness parameter ranges from 1 to 33 for mass accretion rates 
from 0.02 to 0.1, respectively. Combined with the electron temperatures given by the model, we find that the value 
of the compactness parameter is marginally below the critical value for the dominance of electron--positron pair production.
Therefore, the disk corona model is self consistent. Finally, a positive relation between the compactness parameter and photon index is also predicted by the model.

The calculations presented here are restricted to  Eddington ratios up to 0.1 given the upper limit for the accretion flow in a stable ADAF. As such, our models do not address the strong corona inferred from observations of sources at high Eddington ratios. With vertical heat conduction and Compton scattering of strong disk photons,  the upper limit for the accretion rate in a stable corona can differ from that of an ADAF. Investigations under which a strong corona can exist at higher Eddington ratio are planned for the future.

\acknowledgments
Financial support for this work was provided by the National Program on Key Research and Development Project (grant No. 2016YFA0400804) and  the National Natural Science Foundation of China (grant 
11673026).  In addition, R.E.T. acknowledges support 
from the Theoretical Institute for Advanced Research in Astrophysics in the Academia Sinica 
Institute of Astronomy \& Astrophysics.

\bibliographystyle{aasjournal}
\bibliography{compactcorona-v3}

\end{document}